\documentclass[conference]{IEEEtran}
\IEEEoverridecommandlockouts
\usepackage{cite}
\usepackage{amsmath,amssymb,amsfonts}
\usepackage{algorithmic}
\usepackage{graphicx}
\usepackage{textcomp}
\usepackage{multirow}
\usepackage{xcolor}
\usepackage{url}
\usepackage[T1]{fontenc}

\def\BibTeX{{\rm B\kern-.05em{\sc i\kern-.025em b}\kern-.08em
    T\kern-.1667em\lower.7ex\hbox{E}\kern-.125emX}}
\begin{document}

\title{Coronary Artery Disease Classification Using One-dimensional Convolutional Neural Network}
\author{\IEEEauthorblockN{Atitaya Phoemsuk}
\IEEEauthorblockA{\textit{School of Computer Science and Electronic Engineering} \\
\textit{University of Essex}\\
Colchester, United Kingdom \\
ap19698@essex.ac.uk}
\and
\IEEEauthorblockN{Vahid Abolghasemi}
\IEEEauthorblockA{\textit{School of Computer Science and Electronic Engineering} \\
\textit{University of Essex}\\
Colchester, United Kingdom \\
v.abolghasemi@essex.ac.uk}
}

\maketitle

\begin{abstract}
Coronary Artery Disease (CAD) diagnostic to be a major global cause of death, necessitating innovative solutions. Addressing the critical importance of early CAD detection and its impact on the mortality rate, we propose the potential of one-dimensional convolutional neural networks (1D-CNN) to enhance detection accuracy and reduce network complexity. This study goes beyond traditional diagnostic methodologies, leveraging the remarkable ability of 1D-CNN to interpret complex patterns within Electrocardiogram (ECG) signals without depending on feature extraction techniques. We explore the impact of varying sample lengths on model performance and conduct experiments involving layers reduction. The ECG data employed were obtained from the PhysioNet databases, namely the MIMIC III and Fantasia datasets, with respective sampling frequencies of 125 Hz and 250 Hz. The highest accuracy for unseen data obtained with a sample length of 250. These initial findings demonstrate the potential of 1D-CNNs in CAD diagnosis using ECG signals and highlight the sample size's role in achieving high accuracy.
\end{abstract}

\begin{IEEEkeywords}
Convolutional Neural Networks, Electrocardiogram, Coronary Artery Disease, Myocardial Infarction.
\end{IEEEkeywords}

\section{Introduction}
Coronary Artery Disease (CAD) is a significant contributor to mortality worldwide, as reported by the World Health Organisation (WHO) \cite{WHO1}. The condition arises due to the accumulation of plaques, composed of lipid-based substances, which impede the circulation of blood within the arteries. Therefore, timely detection of CAD is imperative, given that the ailment can give rise to debilitating complications such as Congestive Heart Failure (CHF) and Myocardial Infarction (MI), among others. Therefore, it is essential to diagnose an early CAD and prevent death. Numerous studies have stated that no definitive biomarkers or precise electrocardiographic segments can unequivocally indicate the presence of CAD, as different waves and segments are required to be detected \cite{compare_9207044, stackedCNNLSTM}. With the development of artificial intelligence technologies, machine learning and deep learning techniques are being increasingly employed to analyse medical data, encompassing signals, X-rays, Magnetic Resonance Imaging (MRIs), and other modalities. In general medical practices, the electrocardiogram (ECG) is the foremost modality utilised for the preliminary screening of various Cardiovascular Diseases (CVDs). 
Although a recorded ECG may aid in the preliminary diagnosis of CHF and angina, additional tests such as echocardiography and exercise testing are generally required to confirm the diagnosis \cite{medical_study_GPs}. \par
\begin{figure}[h]
    \centering
    \includegraphics[width=0.3\textwidth]{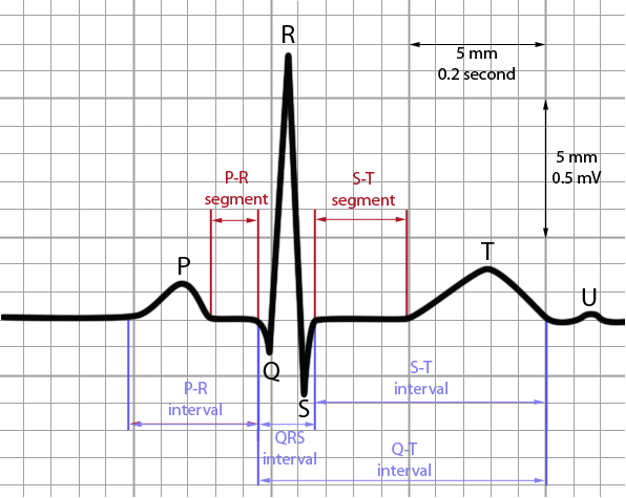}
    \caption{Representation of ECG waveform components\cite{ECGwaves_pic}}
    \label{fig:ecgwaves}
\end{figure}
The morphology of an ECG in a single cycle comprises distinct waveforms, namely, P, Q, R, S, and T waves as shown in Fig. \ref{fig:ecgwaves}. 
ST segment deviation is a vital marker employed in the diagnosis of ischemic conditions, such as CAD, MI, and others. ST depression indicates severe coronary lesions and underscores the significance of an early invasive treatment approach in managing unstable coronary artery disease. Conversely, ST elevation implies complete obstruction of the affected coronary artery and is a hallmark of myocardial infarction or heart attack. In light of recent technological advancements, a multitude of investigators has devised cutting-edge computational diagnosis systems to facilitate the diagnosis of diverse CVDs \cite{Oster_AF_RR, Fan8428414,acharya2017automated}. The analysis of ECG signals for diagnosing CVDs has gained significant attention and has been the focus of an increasing investigation. Numerous deep learning techniques have been employed to classify heart diseases using ECG signals; CNN, Long Short-Term Memory (LSTM) networks, Recurrent Neural Networks (RNNs), and autoencoders \cite{singh2018classification,autoen_9010135}. In recent years, there has been a significant surge in interest among researchers towards using deep learning techniques to diagnose CAD. The majority of researchers have primarily focused on employing CNN techniques for the diagnosis of AF\cite{Fan8428414}, MI \cite{BALOGLU201923}, and arrhythmia\cite{OH2018278}. However, the current research in CAD diagnosis is not yet conclusive due to the limited availability of data and the complex nature of ECG signals in CAD diagnosis. Therefore, a relatively small group of researchers have conducted their work on CAD diagnosis using similar techniques\cite{stackedCNNLSTM,compare_9207044, acharya2017automated}. In \cite{stackedCNNLSTM}, the 1D-CNN was combined with LSTM for CAD diagnosis. It was implemented to extract relevant features from CAD ECG signals. Then, LSTM and a fully connected layer were utilised to conduct the classification. The model is fully automated and needs less feature engineering. However, the limitation of CAD data caused a lower diagnostic performance. Several studies have implemented 1D-CNN for the automated detection of CAD, aiming to enhance diagnostic accuracy and improve patient outcomes. For instance, Acharya et al. \cite{acharya2017automated} proposed an automated CAD diagnosis system based on 1D-CNN, which demonstrated promising results in terms of both accuracy and computational efficiency. 
Feature extraction techniques were combined into a model structure, which later obtained reliable accuracy. However, the model training was time-consuming and needed a large amount of data. Feature extraction still plays an important role in the data pre-processing stage by identifying and selecting informative features in numerous ECG signal processing works 
\cite{acharya2017automated,li2019dual, gupta2021critical,karpagachelvi2010ecg}.\par
In this paper, the primary focus is to investigate the potential applications of a novel and compact 1D-CNN architecture with reduced complexity, with a specific aim of early onset detection. Early detection of CAD is crucial, as it enables timely and suitable treatment, resulting in better health outcomes for patients. To achieve this goal, the proposed 1D-CNN architecture will be developed in order to maintain high performance and minimise computational resource usage. This enhances the model's applicability when performing real-time processing with limited resources.
Additionally, the proposed model will be applied to CAD ECG signals obtained from the MIMIC database. The model's purpose is to capture patterns and distinctive waveform characteristics that serve as markers for the early stages of CAD. To ensure consistent input for our model, a data normalisation technique is implemented to standardise and adjust the ECG signal data, mitigating the impact of noise, variations and artefacts in medical data. By elaborating on these novel aspects of our 1D-CNN architecture, this paper presents a comprehensive and impactful contribution to the field of CAD detection. The model's innovation lies in its efficient architecture, optimising filter counts and kernel size while using dropout layers strategically to enhance early-onset CAD detection precision while conserving computational resources. Through its advancements, the proposed model has the potential to revolutionise early onset CAD detection, ultimately leading to improved patient care and outcomes.
\section{Methodology}
The proposed method consists of three main steps: data collection, pre-processing, and deep learning model. The model will be designed and implemented for CAD classification using the ECG signals through our extensive experiments. Each step will be explained in detail in the following sections.\par

\subsection{Data preparation}
\label{ssec:Data_preparation}
The main portion of ECG data used for training and testing is obtained from the MIMIC III and Fantasia database from the Physionet website\cite{johnson2016mimic_III,Fantasia_database}. A total of approximately 2,840 patients, constituting approximately 7.1\% of all hospital admissions, are identified as having coronary atherosclerosis of the native coronary artery in the MIMIC database. The Fantasia database contains ECGs of 40 healthy patients, including 20 young and 20 adult patients. Three distinct subsets of data are generated for the experiments; $D_{1}$, $D_{2}$ and $D_{3}$. The first subset ($D_{1}$) is created by selecting a cohort of 5 individuals diagnosed with CAD from the MIMIC database, and 5 healthy individuals are chosen from the Fantasia database for the purpose of training and testing the model. The second subset ($D_{2}$) is specifically composed to examine the predictive capabilities of our model further. It comprises 20 CAD subjects from the MIMIC database, alongside 20 non-CAD individuals from the Fantasia database. A third subset ($D_3$) was compiled by selecting patients diagnosed with CAD from the St. Petersburg database \cite{tihonenko2008st}. The St. Petersburg database comprises a total of 7 CAD subjects, with each subject's record spanning a duration of 30 minutes. Each record in the St. Petersburg database consists of 12 standard leads, sampled at a frequency of 257 Hz. The subset $D_2$ and $D_3$ are then utilised for prediction. \par
\subsection{Data pre-processing}
\label{ssec:Data_pre-processing}
\begin{figure*}[!t] 
    \centering
    \includegraphics[width=\textwidth]{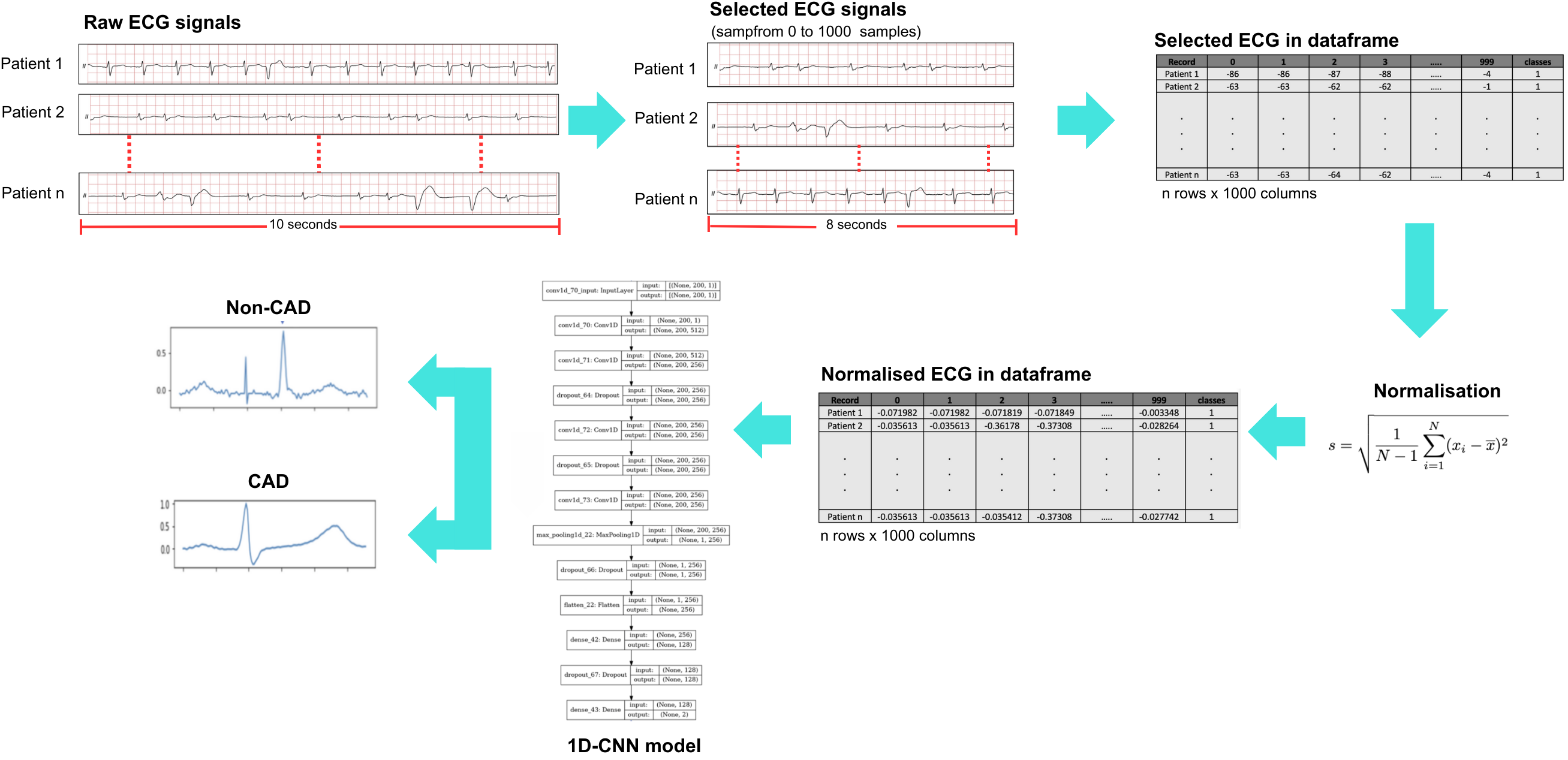}
    \caption{Flowchart of the proposed approach}
    \label{fig:flow_proposed_app}
\end{figure*}

The ECG signals were obtained from patient records, each exhibiting different lengths of signal recordings spanning several minutes. To conduct our experiments effectively, specific segments of these ECG signals were chosen. Initially,  the ECG signal data was retrieved from the records of each patient, as shown in Figure \ref{fig:flow_proposed_app}.  Subsequently, each ECG signal was selected, ranging from 0 to 1000 samples. This segment corresponds to approximately 8 seconds of signal data. The selected data contains a complete cycle of the cardiac waveforms and is then stored in a dataframe.
Prior to inputting into the classifier, the pre-processed data is subjected to labeling. A binary label was assigned to each ECG segment within subsets. Specifically, a label of 0 indicated non-CAD subjects, while a label of 1 indicated CAD subjects. This crucial step is essential for building the basis of supervised learning. The classifier can then acquire valuable features and make informed predictions based on the provided labels. During experiments, the sample lengths were potentially segmented to accommodate the study of the impact of varying lengths on model performance.\par

Data normalisation was then employed to transform numerical data into a standardised range, typically between -1 and 1. This process is achieved by scaling the data based on its mean and standard deviation or by applying a linear transformation to shift and re-scale the data. The standard deviation formula was used for re-scaling, as shown in (1).\par
\begin{equation}
s = \sqrt{\frac{1}{N-1} \sum_{i=1}^N (x_i - \overline{x})^2}
\end{equation}\par
where $s$ is the normalised signal, $N$ is the number of samples, $\overline{x}$ is the average of a given signal, and $x_i$ is the signal value at the $i^{th}$ position in ECG data. The standard deviation measures the spread or dispersion of the signals in the dataset. A smaller standard deviation indicates that the signals are clustered closely around the average, while a larger standard deviation indicates that the signals are more spread out. By re-scaling the data using standard deviation, we obtained normalised data, which was then utilised in classification.\par

\subsection{CNN model}
\label{ssec:CNN_model}
A CNN model consisting of four convolutional layers, a max-pooling layer, four dropout layers, a flattened layer, and a fully connected dense layer was designed. 
The first layer of the network comprises 512 filters with a kernel size of 32, and the subsequent layers contain 256 filters with the same kernel size. By utilising 512 filters with a kernel size of 332 in the initial layer and subsequently reducing the number of filters to 256 while keeping the kernel size consistent, the reduction in parameters contributes to enhancing the model's compactness. The Rectified Linear Unit (ReLU) activation function was used in the convolutional layers to introduce non-linearity into the model. Three dropout layers with a rate of 0.2 were added after the convolutional layers to prevent overfitting. The max pooling layer with a pool size of 128 was then applied to reduce the spatial size of the feature maps and improve generalisation. The flattened output of the max pooling layer was then fed into a fully connected (dense) layer with 128 neurons and ReLU activation, which enabled the model to learn complex representations of the input data. To further prevent overfitting, another dropout layer with a rate of 0.5 was introduced prior to the final output layer. The final output layer consists of two neurons and softmax activation, which enable the model to classify the input data into one of two possible categories. Additionally, to optimise the model's performance, the Adam optimizer with a learning rate of 0.0001 was chosen for parameter optimization. Adam's adaptiveness in adjusting the learning rate for each parameter based on past gradients and magnitudes is particularly beneficial for training CNNs, especially in ECG, where the model must effectively navigate complex, high-dimensional parameter spaces to accurately classify ECG data. The model was compiled using the binary cross-entropy loss function, which is particularly effective for binary classification tasks, such as distinguishing between CAD and non-CAD.

Figure \ref{fig:flow_proposed_app} illustrates the process of classifying ECG signals using the proposed 1D-CNN model. The process starts with raw ECG signals, which are normalised to reduce the impact of variations in amplitude and baseline. The normalised signals are then inputted into the 1D-CNN model, which processes the data and extracts relevant features. The model subsequently uses these features to classify the signals as either CAD or non-CAD. In our study, we proposed a modified 1D-CNN model that included some modifications, such as changes in filter size and the incorporation of dropout layers. This model was employed throughout the CAD analysis process. 
\section{Experimental Results}
\label{sec:Experimental_Results}
In our experiments, we utilised the three subsets prepared during the data preparation stage. The subset $D_{1}$ was split into 70\% for training and 30\% for testing. This split is commonly used in machine learning because it allows for a sufficient amount of data to be used for training, while also providing enough data for testing the model's generalisation ability. Furthermore, the subset $D_{2}$ and $D_{3}$ were used for prediction where the trained model was put to test with these entirely new and unseen subsets.\par

The accuracy was calculated as shown in (2), which measures the proportion of correctly classified instances out of all instances in the dataset. It is a common evaluation metric used to measure the performance of a classification model. 
\begin{equation}
    \text{Accuracy} = \frac{\text{TP} + \text{TN}}{\text{TP} + \text{TN} + \text{FP} + \text{FN}}
\end{equation}\par
Additionally, other critical metrics such as $\text{Misclassification Rate} = \frac{\text{FP} + \text{FN}}{\text{TP + TN + FP + FN}}$, $\text{Precision} = \frac{\text{TP}}{\text{TP} + \text{FP}}$, $\text{Sensitivity} = \frac{\text{TP}}{\text{TP} + \text{FN}}$, and $\text{Specificity} = \frac{\text{TN}}{\text{TN} + \text{FP}}$ were used, where True Positives (TP) are the CAD cases that the model correctly identifies as CAD, True Negatives (TN) are the non-CAD cases correctly identified as non-CAD, False Positives (FP) are the non-CAD cases mistakenly identified as CAD, and False Negatives (FN) are the CAD cases mistakenly identified as non-CAD.
\subsection{Result}
\label{ssec:Resultanddiscussion}
Table \ref{tab:overall} presents the overall performance of the 1D-CNN model in classifying ECG data into CAD and non-CAD categories using different sample lengths, as we aim to determine the optimal sample length for accurate CAD diagnosis using the proposed 1D-CNN model. In the experiment, sample lengths of 1000, 500, 300, 250, 200, and 150 data points were manually selected from the lead II of each ECG subject. Varying the sample length of the input signal can reveal the impact of signal length on the classification model's accuracy. A longer sample length may provide more information about the ECG signal but may also require more sophisticated techniques and longer processing times. On the other hand, a shorter sample length may not be as complex but may lead to lower accuracy due to the loss of critical information in the ECG signal. Hence, identifying the optimal length is essential. \par 
Furthermore, the table shows the results of the experiment conducted on varied lengths of sample size on three different subsets. The model's accuracy in the subset $D_1$ was highest when the sample length was 300 data points, with training accuracy of 100\% and testing accuracy of 96\%, respectively. However, the model's accuracy remained relatively high across all sample lengths for all types of data. Moreover, it indicates that smaller sample lengths generally lead to slightly lower accuracy for train and test data in some sample lengths. The accuracy is significantly increased for unseen data in subset $D_2$ and $D_3$ when smaller sample sizes are employed. The results indicate that the model achieved the highest accuracy for unseen data when the sample length was 250, with an accuracy of 82.5\% in $D_2$ and 85.7\% in $D_3$. \par
\begin{figure}[h]
    \centering
    \includegraphics[width=0.32\textwidth]{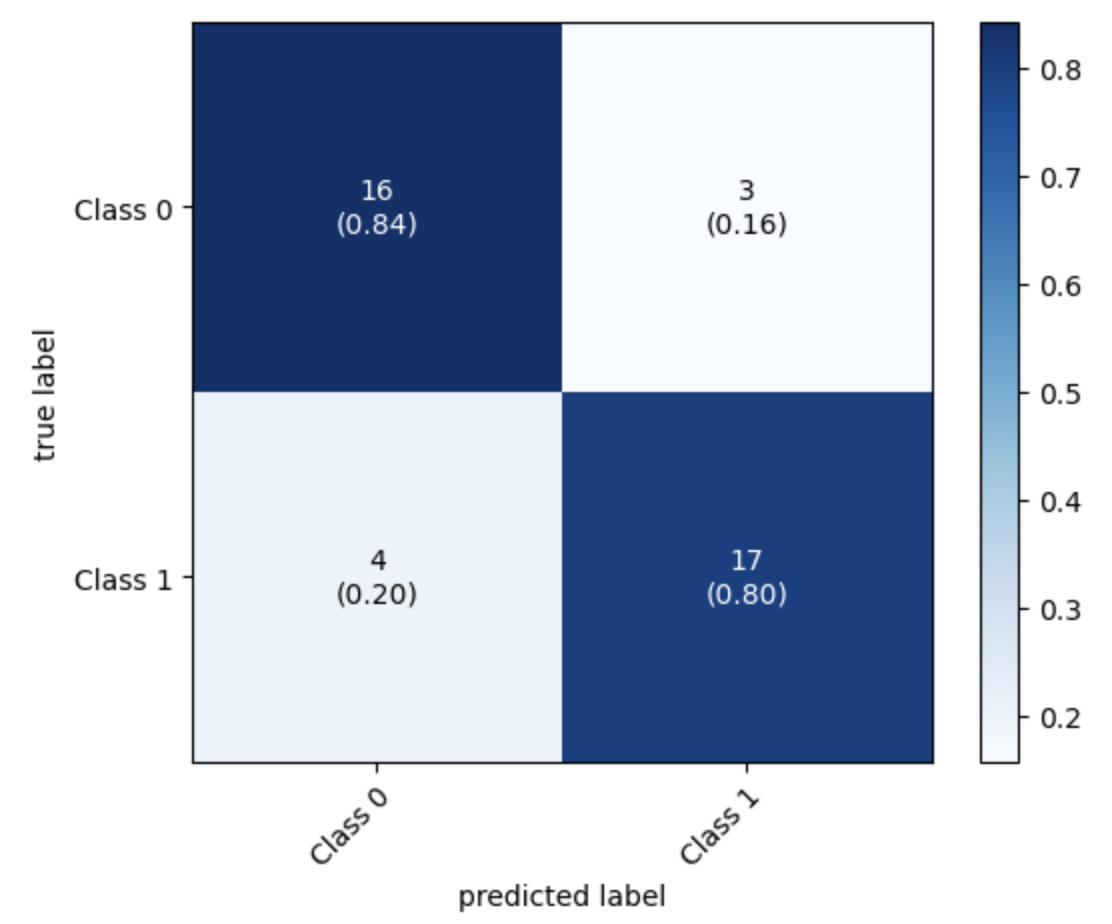}
    \caption{Confusion matrices on dataset $D_2$ with 250-sample length}
    \label{fig:cf}
\end{figure}

Figure \ref{fig:cf} illustrates the performance metrics for CAD detection in $D_2$ and provides valuable insights into the model's effectiveness. With an accuracy of 82.5\%, the model demonstrates its capability to correctly classify all CAD and non-CAD instances, indicating a solid overall performance. However, the misclassification rate of 17.5\% indicates room for improvement in accurately categorising cases. A precision of 85\% represents that when the model identifies a positive case as CAD, it is correct approximately 85\% of the time, showcasing its ability to minimise false positives. A recall of 80\% reflects the model's success in capturing about 80\% of actual CAD cases, which is crucial for avoiding missed diagnoses. Additionally, a specificity of 84\% highlights the model's proficiency in accurately identifying negative cases, implying a satisfactory ability to distinguish non-CAD instances. \par
Overall, the results suggest that a sample length of 250 data points might be optimal for achieving the highest accuracy in subsets $D_2$ and $D_3$ while still maintaining high accuracy for the train and test data in subset $D_1$. This finding could be due to the presence of key features in the ECG signals that indicate CAD, such as ST segments and other important ECG features that may be better represented in a sample length of 250. However, further research is needed to confirm this finding and to explore other factors that might impact the model's performance.\par
\begin{table}[htbp]
\centering
\caption{An overall performance of 1D-CNN on CAD classification using different sample lengths on three subsets.}
\label{tab:overall}
\renewcommand{\arraystretch}{1.5}
\resizebox{0.4\textwidth}{!}{
\begin{tabular}{|c|cccc|}
\hline
\multirow{3}{*}{Sample length} & \multicolumn{4}{c|}{Accuracy (\%)}                                                                                       \\ \cline{2-5} 
                               & \multicolumn{2}{c|}{Subset 1 ($D_1$)}                         & \multicolumn{1}{c|}{Subset 2 ($D_2$)} & \multicolumn{1}{l|}{Subset 3 ($D_3$)} \\ \cline{2-5} 
                               & \multicolumn{1}{c|}{Train} & \multicolumn{1}{c|}{Test} & \multicolumn{1}{c|}{Unseen}    & Unseen                         \\ \hline
150                            & \multicolumn{1}{c|}{95.5}  & \multicolumn{1}{c|}{85}   & \multicolumn{1}{c|}{75.5}      & 81.3                             \\ \hline
200                            & \multicolumn{1}{c|}{94.6}  & \multicolumn{1}{c|}{90}   & \multicolumn{1}{c|}{73.8}      & 85.6                           \\ \hline
250                            & \multicolumn{1}{c|}{97.3}  & \multicolumn{1}{c|}{89}   & \multicolumn{1}{c|}{82.5}        & 85.7                           \\ \hline
300                            & \multicolumn{1}{c|}{100}   & \multicolumn{1}{c|}{96}   & \multicolumn{1}{c|}{71.9}      & 82.4                           \\ \hline
500                            & \multicolumn{1}{c|}{98}    & \multicolumn{1}{c|}{82}   & \multicolumn{1}{c|}{66.7}      & 63                             \\ \hline
1000                           & \multicolumn{1}{c|}{95.6}  & \multicolumn{1}{c|}{89}   & \multicolumn{1}{c|}{63.6}      & 65                             \\ \hline
\end{tabular}
}
\end{table}

Table \ref{tab:comparsion_CNN} provides a comparative overview of the performance of three distinct approaches for CAD detection using the MIMIC III dataset. The Baseline 1D-CNN approach demonstrates a reasonable level of performance, achieving an accuracy of 83\% on the training set. This result indicates its capacity to learn from the training data and identify patterns associated with CAD. However, a noteworthy observation is the decrease in accuracy to 74\% on the test set. The complexity of the model could lead to the capturing of irrelevant features during training, resulting in a noticeable decrease in testing accuracy. The Hybrid CNN-LSTM approach enhances the ability to identify relevant CAD features by introducing LSTM layers. This model achieved an accuracy of 94\% in both the training and test sets, indicating effective generalisation and feature extraction capabilities. Lastly, our proposed model demonstrated remarkable accuracy, achieving 97.3\% on the training set and 89\% on the test set. However, the noticeable drop in accuracy on the test set warrants further exploration and investigation.\par

\begin{table}[]
\centering
\renewcommand{\arraystretch}{1.5}
\caption{Comparison of existing approaches performance on CAD applications using the MIMIC III dataset.}
\label{tab:comparsion_CNN}
\tiny
\resizebox{0.38\textwidth}{!}{
\begin{tabular}{|c|cl|}
\hline
\multirow{2}{*}{Architecture}                  & \multicolumn{2}{c|}{Accuracy (\%)}                             \\ \cline{2-3} 
                                               & \multicolumn{1}{c|}{Train set} & \multicolumn{1}{c|}{Test set} \\ \hline
\multicolumn{1}{|l|}{Baseline 1D-CNN\cite{compare_9207044}} & \multicolumn{1}{c|}{83} &  \multicolumn{1}{c|}{74}    \\ \hline
\multicolumn{1}{|l|}{Hybrid CNN-LSTM\cite{compare_9207044}} & \multicolumn{1}{c|}{94} &  \multicolumn{1}{c|}{94}    \\ \hline
\multicolumn{1}{|l|}{\textbf{Proposed model}} & \multicolumn{1}{c|}{97.3} &  \multicolumn{1}{c|}{89}    \\ \hline
\end{tabular}
}
\end{table}

\begin{table}
\caption{Comparison of dropout layer configurations and probabilities in the proposed model}
\centering
\resizebox{0.48\textwidth}{!}{
\renewcommand{\arraystretch}{1.5}
\begin{tabular}{|l|ll|}
\hline
\multicolumn{1}{|c|}{\multirow{2}{*}{Architecture}} & \multicolumn{2}{c|}{Accuracy(\%)}                                   \\ \cline{2-3} 
\multicolumn{1}{|c|}{}                                                 & \multicolumn{1}{c|}{Unseen ($D_{2}$)} & \multicolumn{1}{c|}{Unseen ($D_{3}$)} \\ \hline
No dropout layers                                                      & \multicolumn{1}{l|}{62}          & 60                               \\ \hline
one dropout (0.2)                                                      & \multicolumn{1}{l|}{68}          & 71                               \\ \hline
two dropout (0.2)                                                      & \multicolumn{1}{l|}{65}          & 57                               \\ \hline
three dropout (0.2)                                                    & \multicolumn{1}{l|}{65}          & 62                               \\ \hline
three dropout (0.2) and a dropout of (0.5)                            & \multicolumn{1}{l|}{79}          & 86                               \\ \hline
\end{tabular}
}
\label{tab:comparsion_dropout}
\end{table}

In addition, an ablation experiment encompassing diverse configurations of dropout layers was conducted to evaluate the efficacy of the integration of these layers on the optimal sample length, as shown in Table \ref{tab:comparsion_dropout}. The results of these experiments reveal a significant improvement in the performance of subset $D_{2}$, achieved through the incorporation of four dropout layers, each configured with dropout rates of 0.2 and 0.5. Notably, this configuration achieves the highest accuracy of 79\% in $D_{2}$ and 86\% in $D_{3}$. This improvement strongly suggests that adding dropout layers helps enhance the model's ability to generalise and effectively addresses concerns about overfitting.  However, it is important to exercise caution when considering the inclusion of more dropout layers or higher dropout rates, as these adjustments may not necessarily lead to further performance gains. In fact, excessive dropout can potentially hinder the network's learning capacity. The experimental results demonstrate the effectiveness of the 1D-CNN model in accurately classifying ECG data into CAD and non-CAD categories, regardless of the sample length. The model achieved high accuracy for both train and test data across all sample lengths, with the highest accuracy observed when the sample length was set to 300 data points. This indicates that the model was able to learn and generalise well from various ECG samples, regardless of their length. Interestingly, the experimental findings suggest that reducing the sample length leads to a slight decrease in the accuracy of both train and test data. However, this is compensated by a significant improvement in the accuracy of unseen data, highlighting the potential for better generalisation of the 1D-CNN model with smaller sample sizes.\par

\begin{table*}[t]
\centering
\caption{Comparative analysis of model complexity with existing work}
\renewcommand{\arraystretch}{1.5}
\label{tab:comparison_complexity}
\begin{tabular}{|l|c|c|c|}
\hline
Metric & Proposed Model & Baseline 1D-CNN\cite{compare_9207044} & Hybrid CNN-LSTM\cite{compare_9207044} \\ \hline
Number of Layers & 12 & 14 & 14 \\ \hline
No. of Parameters & 8 M & 0.4 M & 4 M \\ \hline
Activation Function & ReLU & ReLU & ReLU \\ \hline
Pooling Layers & Max Pooling & Max Pooling & Max Pooling \\ \hline
Dropout Rate & 0.2 and 0.5 & 0.2 & 0.2 \\ \hline
Learning Rate & 0.001 & 0.003 &  0.003 \\ \hline
Floating-Point Operations (FLOPs) & 65,792 & 71,936 & 23,609,344 \\ \hline
Performance (Accuracy) & 97.3\% & 83\% & 94\% \\ \hline
Computational Resources & Apple M2 Max & Intel® Xeon(R) 16-core & Intel® Xeon(R) 16-core \\ \hline
\end{tabular}
\end{table*}

Additionally, we conducted an extensive examination of the model's complexity, as illustrated in Table \ref{tab:comparison_complexity}. The Baseline 1D-CNN and the Hybrid CNN-LSTM models, each comprising 14 layers, exhibit significant differences in parameter usage. The Baseline 1D-CNN employs 0.4 million parameters, while the Hybrid CNN-LSTM utilises 4 million parameters and achieves 83\% and 94\% accuracy, respectively. However, the proposed model features a relatively complex architecture with 12 layers and 8 million parameters. Remarkably, despite its relatively lower complexity and the smaller dataset size, it attains an impressive accuracy rate of 97.3\%. The proposed model outperforms the Baseline 1D-CNN and Hybrid CNN-LSTM in accuracy but requires significantly more computational resources, limiting its practicality in resource-constrained environments.\par
Furthermore, the proposed model demonstrates remarkable efficiency in terms of computational complexity, as evidenced by its significantly lower demand for Floating-Point Operations (FLOPs) compared to the Hybrid CNN-LSTM and Baseline 1D-CNN approaches. With only 65,792 FLOPs, our proposed model achieves outstanding performance, surpassing the accuracy of the Baseline 1D-CNN and Hybrid CNN-LSTM. This efficiency translates into a more cost-effective and energy-efficient deployment, making it an attractive option for real-world applications. However, it is important to note that a direct correlation between a number of parameters and FLOPs is not definitively established. Increasing these factors does not consistently lead to better performance. In the proposed model, it is noteworthy that it exhibits enhanced performance while requiring fewer computational resources despite having the highest number of parameters and the lowest number of FLOPs.\par To summarise, the proposed model emerges as a suitable choice, distinguished by its exceptional computational efficiency, high accuracy, and resource-efficient design in contrast to the remaining models in the table. It successfully balances complexity and performance, offering a practical and cost-effective solution for real-world applications.

\section{Conclusion}
\label{Conclusion}
Given the absence of precise CAD biomarkers, identifying robust classification features becomes crucial. Exploring alternative ECG channels for CAD detection is also deemed essential. Extracting CAD-specific data from diverse patient records in the MIMIC III database proved challenging due to the varied ECG storage methods across channels. Future research should persist in exploring feature extraction techniques and their impact on model performance while considering their limitations. Furthermore, upcoming studies could delve into other critical factors influencing the model's performance, such as the number of ECG leads utilised, sample size, and additional underlying medical conditions. One of the most pivotal aspects to further address is reducing network complexity, as it is directly correlated with model accuracy.

\bibliography{ref}
\bibliographystyle{plain} 

\end{document}